\documentclass[12pt,a4paper]{article}
\pdfoutput=1
\usepackage[a4paper]{geometry}
\usepackage{graphicx}
\usepackage{amsmath}
\usepackage{amssymb}
\usepackage{float}
\usepackage{listings}
\title{\bf Probing the BFKL dynamics in inclusive three jet production at the LHC}

\author{F. Caporale$^1$, F.~G. Celiberto$^{1,2}$, G. Chachamis$^1$, \\ 
        D. Gordo G{\' o}mez$^1$\footnote{`la Caixa'-Severo Ochoa Scholar.}\,\,,  A. Sabio Vera$^{1}$\\ \\
{\small $^1$ Instituto de F{\' \i}sica Te{\' o}rica UAM/CSIC, Nicol{\'a}s Cabrera 15}\\ 
{\small \& Universidad Aut{\' o}noma de Madrid, E-28049 Madrid, Spain.}\\
{\small $^2$ Dipartimento di Fisica, Universit{\`a} della Calabria \&}\\
{\small Istituto Nazionale di Fisica Nucleare, Gruppo Collegato di Cosenza,}\\
{\small I-87036 Arcavacata di Rende, Cosenza, Italy.}
}

\begin{document}

\maketitle 

\abstract
We propose the study of new observables in LHC inclusive events with three tagged jets, one in the forward direction, one in the backward direction and both well-separated in rapidity from the each other (Mueller-Navelet jets), together with a third jet tagged in central regions of rapidity. Since non-tagged associated mini-jet multiplicity is allowed, we argue that projecting the cross sections on azimuthal-angle components can provide several distinct tests of the BFKL dynamics. Realistic LHC kinematical cuts are introduced.

\section{Introduction}
\label{intro}

The Balitsky-Fadin-Kuraev-Lipatov (BFKL) resummation program in the leading logarithmic (LL)~\cite{Lipatov:1985uk,Balitsky:1978ic,Kuraev:1977fs,Kuraev:1976ge,Lipatov:1976zz,Fadin:1975cb} and next-to-leading logarithmic (NLL) approximation~\cite{Fadin:1998py,Ciafaloni:1998gs} may be applied for phenomenological studies  at hadronic colliders when the final-state is characterised by jets that are produced at large relative rapidities. 
Mueller-Navelet jets~\cite{Mueller:1986ey} is an key example, specifically, for observables that are based on
the azimuthal angle formed by the two outermost in rapidity tagged jets, $\phi$. The precise form of the
observables is built by considering
ratios of projections on the azimuthal angle~${\cal R}^m_n = \langle \cos{(m \, \phi)} \rangle / \langle \cos{(n \, \phi)} \rangle$. 
Comparison of different NLL predictions against LHC experimental data for these observables has  been quite successful
~\cite{DelDuca:1993mn,Stirling:1994he,Orr:1997im,Kwiecinski:2001nh,Angioni:2011wj,Caporale:2013uva, Caporale:2013sc,Marquet:2007xx,Colferai:2010wu,Ducloue:2013wmi,Ducloue:2014koa,Mueller:2015ael,N.Cartiglia:2015gve,Chachamis:2015crx,Vera:2006un,Vera:2007kn,Ducloue:2013bva,Caporale:2014gpa,Caporale:2014blm,Celiberto:2015dgl,Celiberto:2016ygs}.

\begin{figure}[h]
\centering
     \includegraphics[scale=0.35]{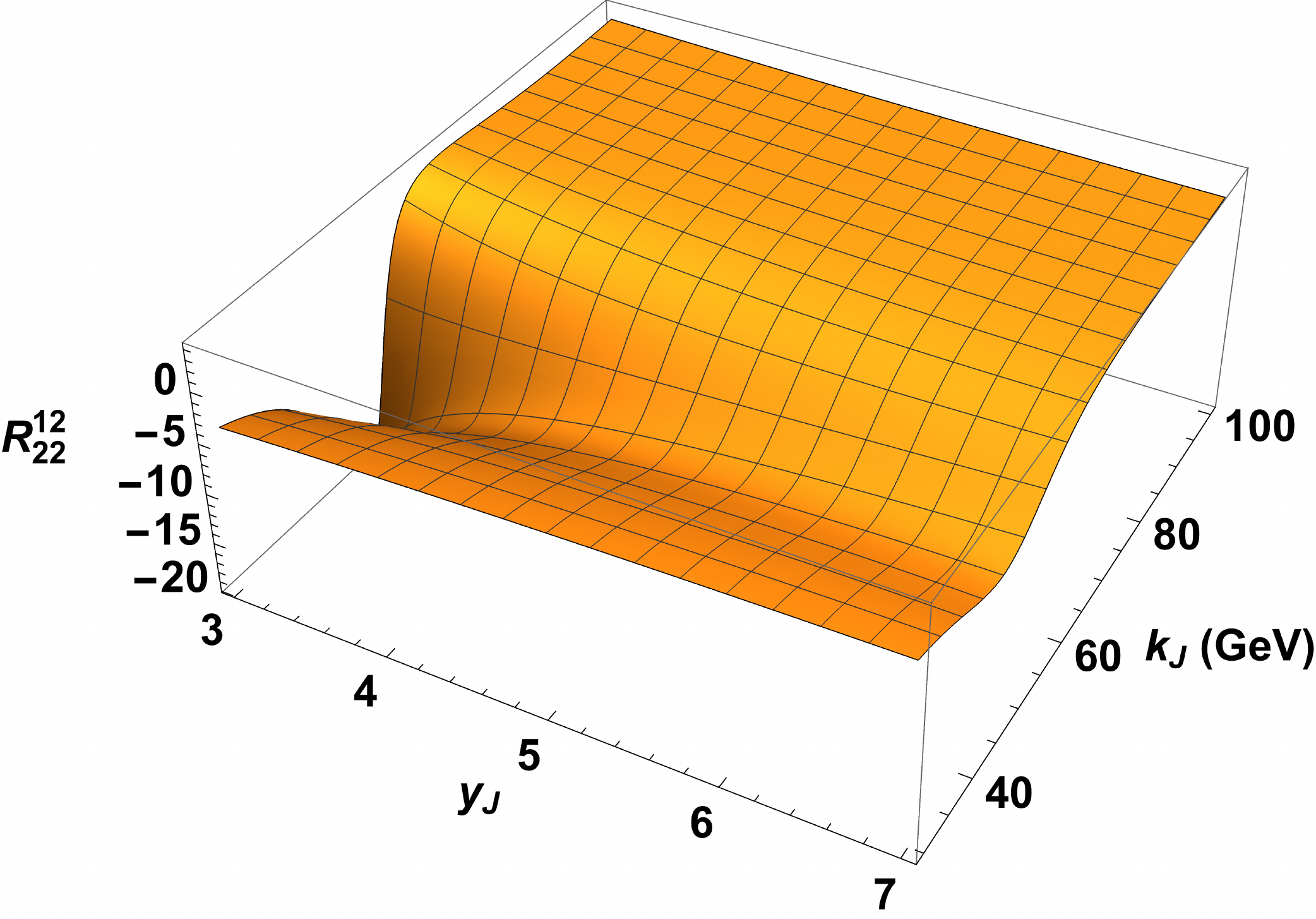}
\caption{\small 3D plot of the partonic ${\cal R}^{12}_{22}$ as a function 
of  the rapidity $\text{y}_J$ and the momentum $k_J$ of the 
central jet for $k_A = 40$ GeV, $k_B = 50$ GeV and $\Delta Y_{A,B}=10$.} 
\label{3d}
\end{figure}
New LHC observables, that may be seen as a generalisation
of the Mueller-Navelet jets, were proposed recently for inclusive three-jet~\cite{Caporale:2015vya,Caporale:2016soq}
and four-jet production~\cite{Caporale:2015int,Caporale:2016xku}. In this work we discuss only the
observables for inclusive three-jet production.
These are defined by the generalised ratios~\cite{Caporale:2015vya}
\begin{eqnarray}
{\cal R}^{M N}_{P Q} =\frac{ \langle \cos{(M \, \phi_1)} \cos{(N \, \phi_2)} \rangle}{\langle \cos{(P \, \phi_1)} \cos{(Q \, \phi_2)} \rangle} \, , 
\label{Rmnpq}
\end{eqnarray}
where $\phi_1$ is the azimuthal angle difference between the forward and the central jet
and $\phi_2$  the azimuthal angle difference between the central jet and the backward in
rapidity jet.
The ratios ${\cal R}^{M N}_{P Q}$ in Eq.~(\ref{Rmnpq}) are actually partonic level 
quantities and therefore, cannot be readily compared to experimental data. 
Therefore, we define the hadronic level
observables $R^{M N}_{P Q}$~\cite{Caporale:2016soq} and study their stability once
we introduce corrections beyond the LL accuracy.
For that, we produce the two outermost in rapidity jets
 within the collinear factorization scheme,  each 
 of them associated with a forward ``jet vertex''~\cite{Caporale:2012:IF}. Then we link
  these vertices and the 
 central jet using two BFKL gluon Green's functions. At the end, the partonic differential
 cross-section is convoluted with collinear parton distribution functions
 and is integrated over the momenta of all produced jets
in order to calculate the ratios $R^{M N}_{P Q}$.  For the integration
 over the momenta of the jets we use standard LHC experimental cuts. 
The rapidity of the central jet takes values close to the middle of the
rapidity distance between the two outermost tagged jets.
 
\section{Hadronic inclusive three-jet production in multi-Regge kinematics}
\label{sec-1}

Let us first remember some of the notation defined in~\cite{Caporale:2015vya,Caporale:2016soq}.
Assuming that the transverse momenta of the outermost jets are $\vec{k}_{A,B}$, their rapidity difference, $Y$, is large and the central jet has transverse momentum $\vec{k}_J$. We allow for mini-jet activity  between the three tagged jets so that the process\footnote{Another interesting idea,  suggested in~\cite{Ivanov:2012iv} 
and investigated in~\cite{Celiberto:2016hae,celiberto:wip}, is
the study of the production of two charged light hadrons,
$\pi^{\pm}$, $K^{\pm}$, $p$, $\bar p$, with large transverse momenta and well
separated in rapidity.} we need to study is
\begin{eqnarray}
\label{process}
{\rm proton }(p_1) + {\rm proton} (p_2) \to  
{\rm jet}(k_A) + {\rm jet}(k_J) + {\rm jet}(k_B)  + {\rm minijets}\;.
\end{eqnarray}

Firstly, we define
the two relative azimuthal angles between the outermost jets
and the central jet as
$\Delta\theta_{\widehat{AJ}} = \theta_A - \theta_J - \pi$ and 
$\Delta\theta_{\widehat{JB}} = \theta_J - \theta_B - \pi$ respectively.
Then the projection on azimuthal-angle components  gives
the mean value 
\begin{eqnarray}
\label{Cmn}
 {\cal C}_{MN} \, = \,
 \langle \cos{\left(M \left( \theta_A - \theta_J - \pi\right)\right)}  
 \cos{\left(N \left( \theta_J - \theta_B - \pi\right)\right)}
 \rangle && \\
 &&\hspace{-9cm} = \frac{\int_0^{2 \pi} d \theta_A d \theta_B d \theta_J \cos{\left(M \left( \theta_A - \theta_J - \pi\right)\right)}  \cos{\left(N \left( \theta_J - \theta_B - \pi\right)\right)}
 d\sigma^{3-{\rm jet}} }{\int_0^{2 \pi} d \theta_A d \theta_B d \theta_J 
 d\sigma^{3-{\rm jet}} },\nonumber
\end{eqnarray}
where $M,N$ are positive integers)
and  $d\sigma^{3-{\rm jet}}$ the partonic 
deferential cross-section for three-jet production defined in ~\cite{Caporale:2015vya}.

\begin{figure}[h]
\centering
   \includegraphics[scale=0.45]{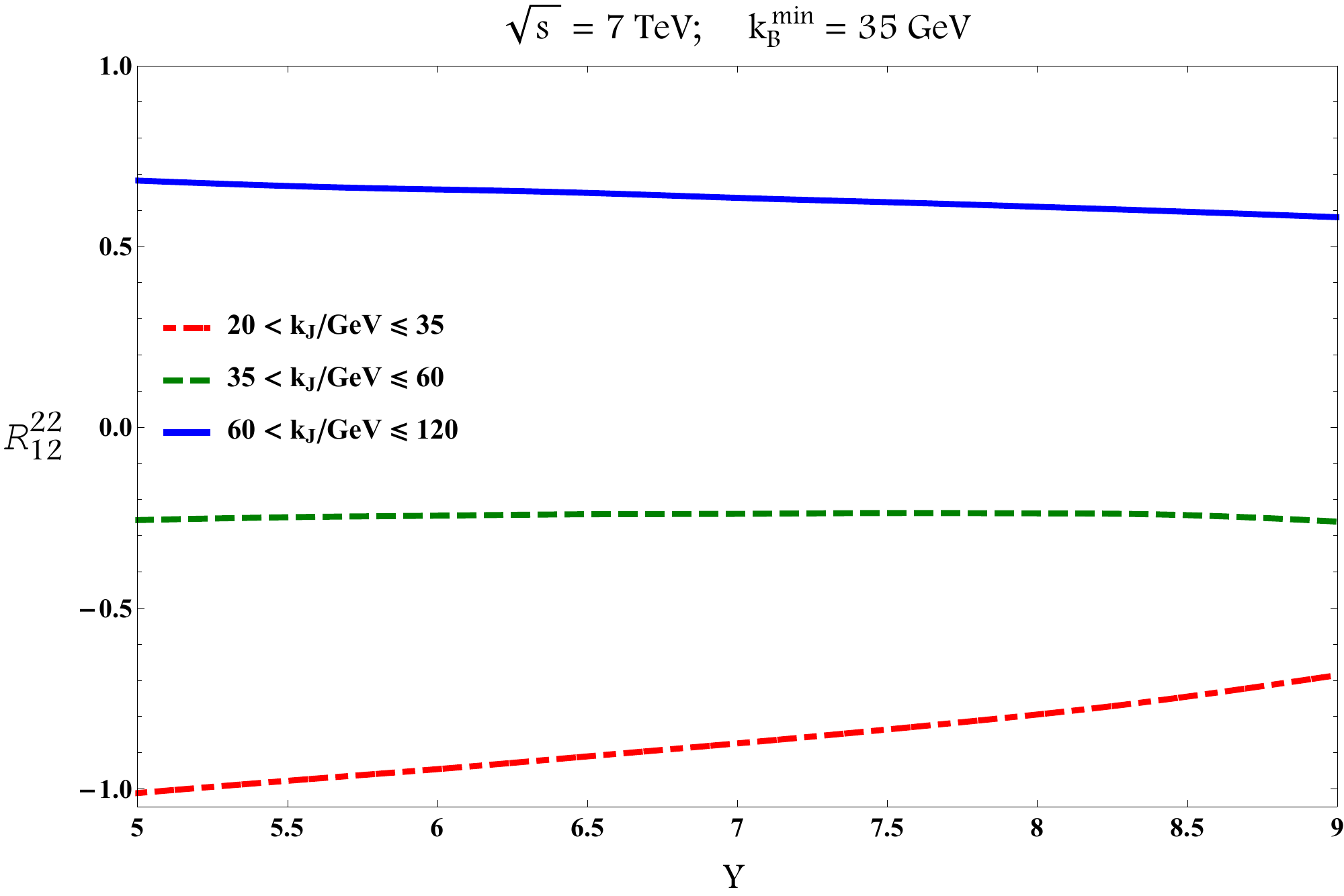}
   \includegraphics[scale=0.45]{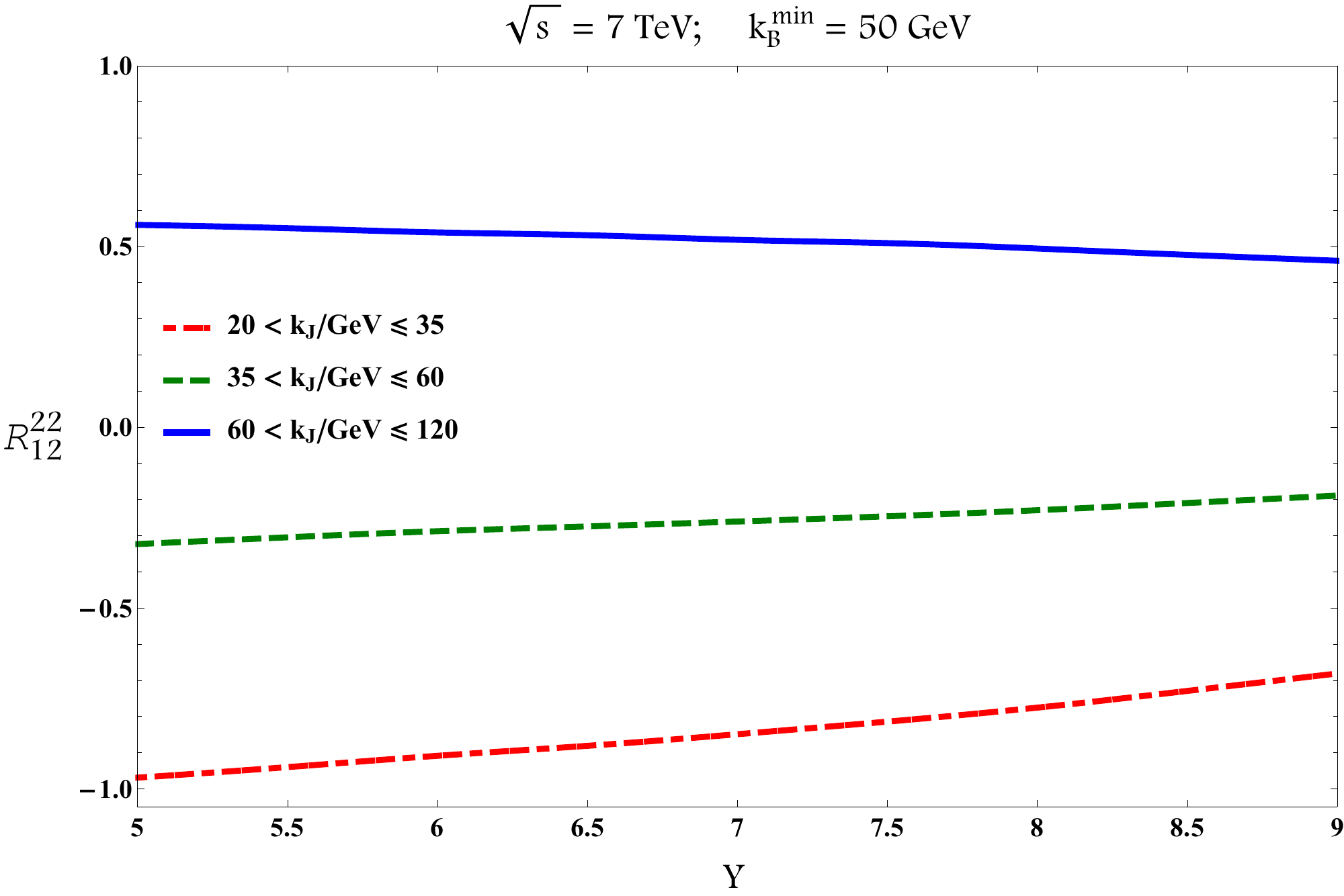}
\caption{\small $Y$-dependence of the LL
$R^{22}_{12}$ for $\sqrt s = 7$ TeV.
Symmetric cut $k_B^{\rm min} = 35$ GeV (top) and asymmetric cut $k_B^{\rm min} = 50$ GeV (bottom).} 
\label{res1}
\end{figure}

In order to compute theoretical estimates that may be compared against
current and future experimental data, 
we integrate ${\cal C}_{M,N}$ over the momenta of the tagged jets in the form
\begin{align}
\label{Cmn_int}
 C_{MN} =
 \int_{Y_A^{\rm min}}^{Y_A^{\rm max}} \hspace{-0.25cm} dY_A
 \int_{Y_B^{\rm min}}^{Y_B^{\rm max}} \hspace{-0.25cm} dY_B
 \int_{k_A^{\rm min}}^{k_A^{\rm max}} \hspace{-0.25cm} dk_A
 \int_{k_B^{\rm min}}^{k_B^{\rm max}} \hspace{-0.25cm} dk_B
 \int_{k_J^{\rm min}}^{k_J^{\rm max}} \hspace{-0.25cm} dk_J
 \delta\left(Y_A - Y_B - Y\right) {\cal C}_{MN},
\end{align}
where the rapidity of the forward jet takes values in the
range  $Y_{A}^{\rm min}  = 0$  and 
$Y_{A}^{\rm max}  = 4.7$ and that
of the backward jet in the
range  $Y_{B}^{\rm min}  = -4.7$  and 
$Y_{B}^{\rm max}  = 0$ while their difference 
$Y \equiv Y_A - Y_B$ is kept fixed at definite values in the range $5 < Y < 9$.
We calculate $C_{MN}$ for two different 
center-of-mass energies, $\sqrt s = 7$ and $\sqrt s = 13$ TeV and
we introduce two typical kinematical cuts  previously
used in the study of Mueller-Navelet jets at the LHC. Specifically,
we use both a symmetric and an asymmetric
cut~\cite{Ducloue:2013wmi,Celiberto:2015dgl}:
\begin{enumerate}
\item $k_A^{\rm min} = 35$ GeV, $k_B^{\rm min} = 35$ GeV, $k_A^{\rm max} = k_B^{\rm max}  = 60$ GeV
(symmetric); \,
\item $k_A^{\rm min} = 35$ GeV, $k_B^{\rm min} = 50$ GeV,  $k_A^{\rm max} = k_B^{\rm max}  = 60$ GeV
(asymmetric).
\end{enumerate}
\begin{figure}[h]
\centering
   \includegraphics[scale=0.45]{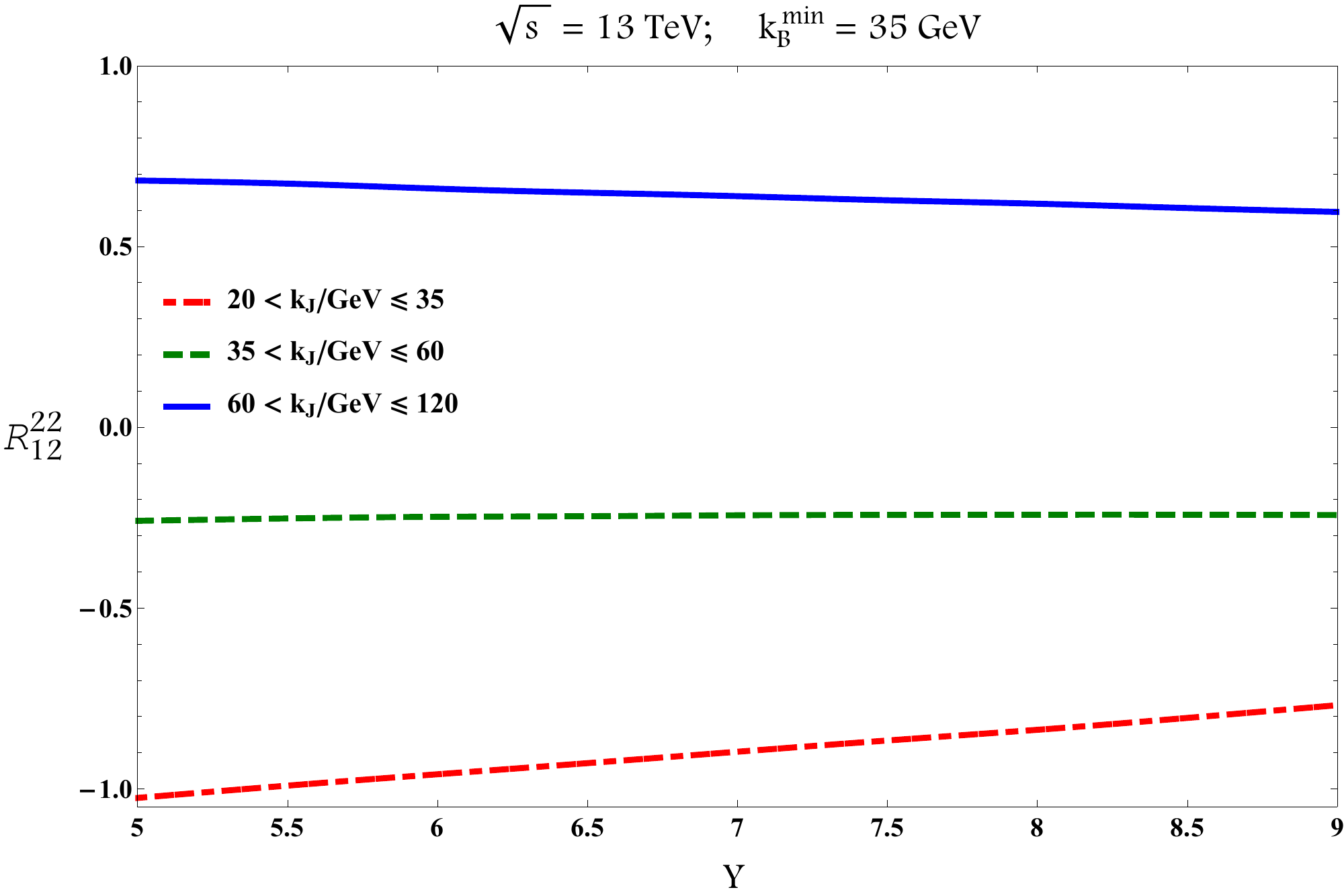}
   \includegraphics[scale=0.45]{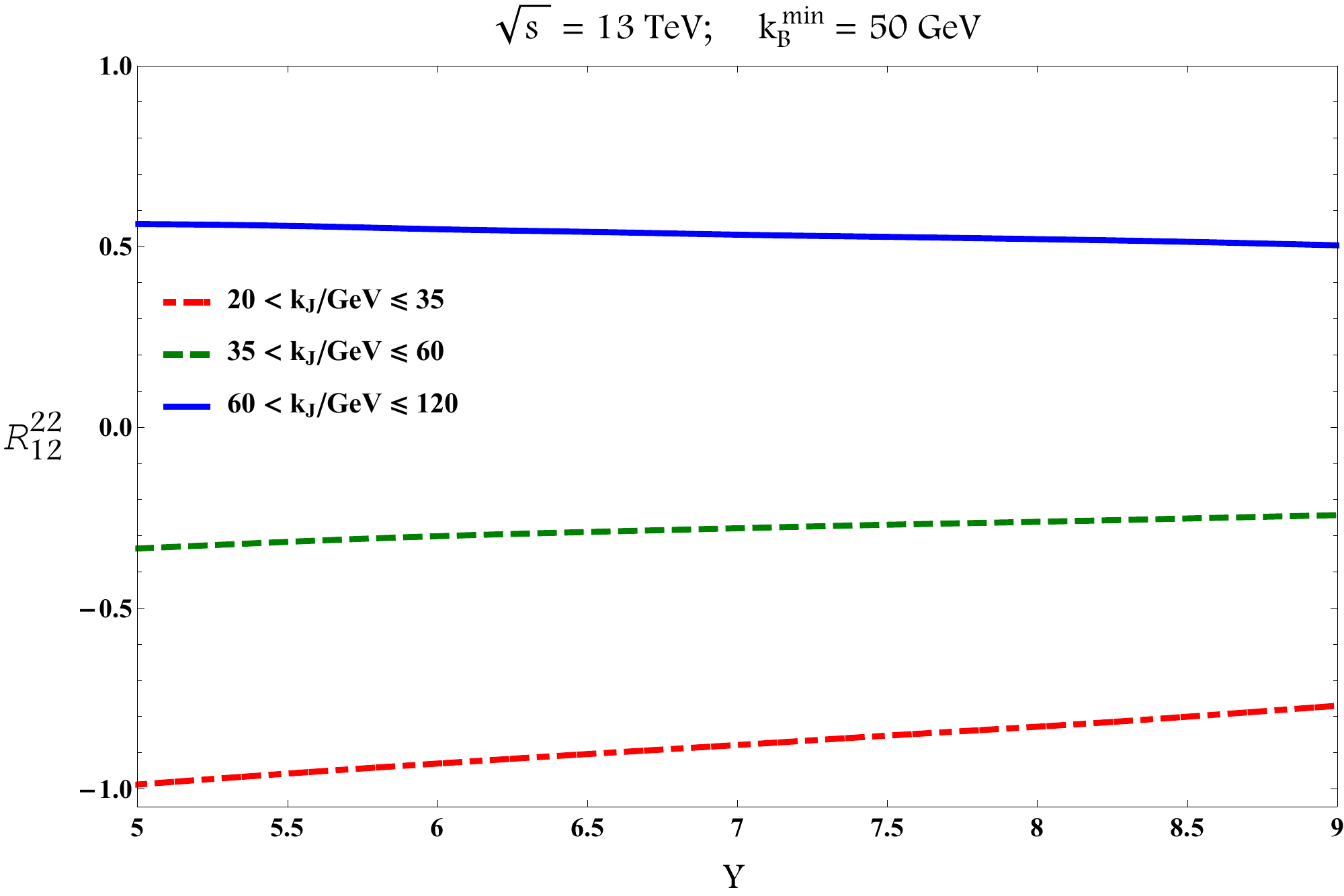}
\caption{\small $Y$-dependence of the LL
$R^{22}_{12}$ for $\sqrt s = 13$ TeV.
Symmetric cut $k_B^{\rm min} = 35$ GeV (top) and asymmetric cut $k_B^{\rm min} = 50$ GeV (bottom).} 
\label{res2}
\end{figure}
We are interested in maximising the 
stability with respect to higher order effects (beyond LL)  in our 
results (see~\cite{Caporale:2013uva}), therefore, 
we remove the  
zeroth conformal spin contribution of the BFKL kernel
by considering the ratios
\begin{eqnarray}
\label{RPQMN}
R_{PQ}^{MN} \, = \, \frac{C_{MN}}{C_{PQ}}, \,\,\,\,\,\,\,\, M, N, P, \,Q > 0,
\label{RmnqpNew}
\end{eqnarray}
which have no $n=0$ dependence. Thus, 
we can study the ratios  $R_{PQ}^{MN}(Y)$ in Eq.~(\ref{RmnqpNew}) as functions of the 
rapidity difference $Y$ between the outermost jets 
for some typical values of $M, N, P, Q$.
We define three different
$p_T$ ranges (bins) for the allowed momentum of the central jet: \\
$\text{bin-1} = [20\, \mathrm{GeV} < k_J < 35\, \mathrm{GeV}]$  ($k_J$ `smaller' that $k_A$, $k_B$),\\
$\text{bin-2} = [35 \,\mathrm{GeV} < k_J < 60\, \mathrm{GeV}]$ ($k_J$ `similar' to $k_A$, $k_B$), \\
$\text{bin-3} = [60\, \mathrm{GeV} < k_J < 120\, \mathrm{GeV}]$ ($k_J$ `larger' than $k_A$, $k_B$).\\
 This
 permits the discrimination of different behaviours of the $R_{PQ}^{MN}(Y)$
 by using as a criterion the relative size of the central jet. 
 In Fig.~\ref{3d} we show the behaviour of ${\cal R}^{12}_{22}$ as we change the size of the central
jet and its position in rapidity. We notice that while a small variation in $\text{y}_J$ around the central
rapidity value $\Delta Y_{A,B}/2=5$ does not result in significant changes for a fixed $k_J$, a change in the value
of $k_J$ may have a big impact for a fixed $\text{y}_J$. 
\begin{figure}[h]
\centering
   \includegraphics[scale=0.45]{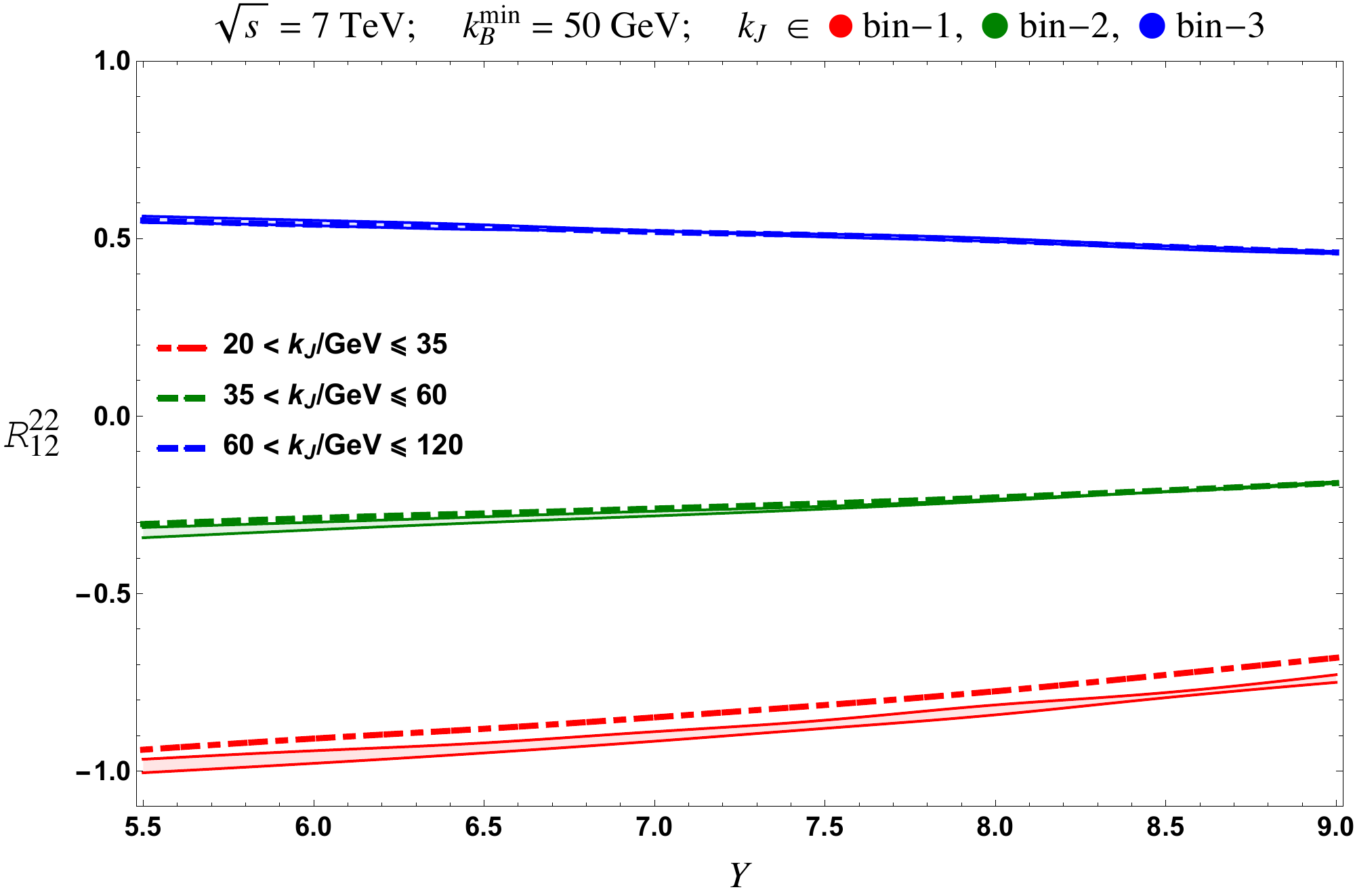}
   \includegraphics[scale=0.45]{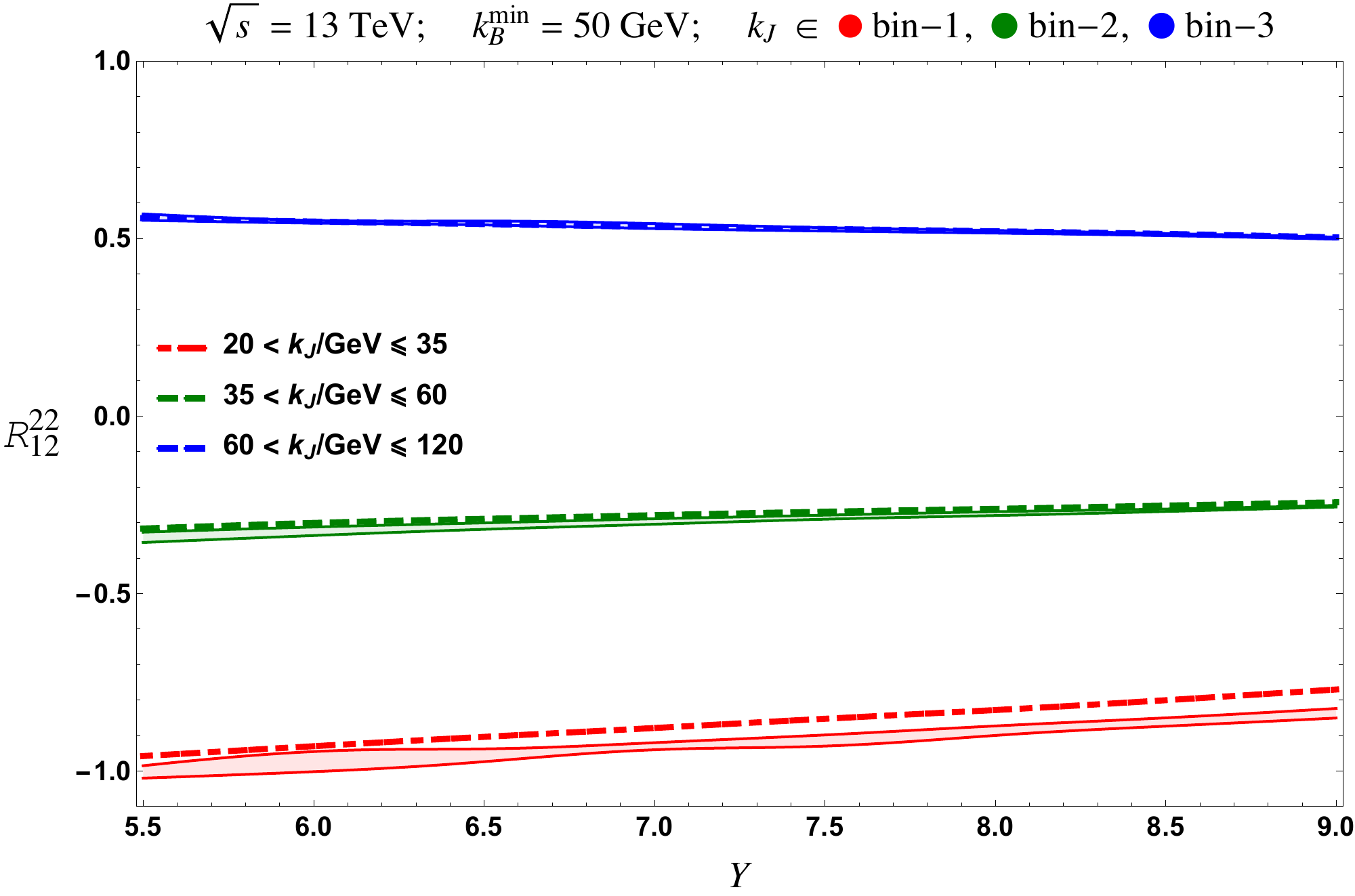}
\caption{\small 
$Y$-dependence of the LL (dashed lines) and NLL (BLM) predictions (continuous bands) for
$R^{22}_{12}$ in the asymmetric cut at $\sqrt s = 7$ TeV (top) and
$\sqrt s = 13$ TeV (bottom). The blue NLL band is very narrow and lies on top of the LL line. } 
\label{res3}
\end{figure}
A number of different ratios was presented in~\cite{Caporale:2016soq}, 
here we are focussing on ratios that involve the coefficients
$C_{12}$ and $C_{22}$.
In Figs.~\ref{res1} and~\ref{res2} we see the LL accuracy results for  $R^{22}_{12}$.
Generally, the dependence of the different observables on the rapidity
difference between $k_A$ and $k_B$ is rather smooth whereas 
the slope of the three curves depends
on the particular observable. 
For $R^{22}_{12}$ we see that
shifting from a symmetric to an asymmetric cut makes no noticeable difference.
Moreover, there are no important changes when we 
change the colliding energy from $\sqrt{s} = 7$ TeV to $\sqrt{s} = 13$ TeV.
The latter is crucial as it suggests that $R^{22}_{12}$ is already within
some sort of asymptotic regime
for the specific kinematical configurations. 

Apart from the stability of the observable with regard to an increase of the colliding energy,
another important question is the stability with respect to effects that go beyond the
LL approximation~\cite{NLO-3jets}. A first important step towards a full NLL computation is to take into account
the NLL contributions to the two gluon Green's functions that connect the three jets. In Fig.~\ref{res3}
we present exactly these corrections obtained by using
 the Brodsky-Lepage-Mackenzie (BLM) prescription~\cite{Brodsky:1982gc}
for  the $R^{22}_{12}$ coefficient in the asymmetric cut. 
In particular, we have used the MOM scheme and chosen the renormalisation scale such that the
$\beta_0$-dependence of the given observable vanishes, following the BLM prescription.
 The dashed lines represent  the LL predictions and the coloured bands represent the NLL BLM predictions.
 It is impressive that the NLL values are almost on top of the LL ones which gives us great confidence
that the observables $R^{MN}_{PQ}$ are indeed excellent BFKL probes at the LHC.

\section{Summary \& Outlook}

We have presented a first BFKL driven hadronic-level phenomenological work 
on the recently proposed observables $R^{MN}_{PQ}$ that depend on the azimuthal-angles of the jets in
inclusive three-jet production at the LHC.
In particular, we concentrated on the ratio $R^{22}_{12}$.
Our major task was to study the variation of $R^{22}_{12}$
when we set the rapidity
difference $Y$ between the outermost jets to different fixed values
in the range $5 < Y < 9$.
Generally, we notice a smooth functional dependence of
the ratio $R^{22}_{12}$ on $Y$. A key observation is that $R^{22}_{12}$
and  other similar ratio observables do not exhibit a significantly different behaviour
when one changes the energy configuration from 7 to 13 TeV. This gives us confidence that
these ratios pinpoint the most important characteristics of the azimuthal behaviour
of the tagged jets within the BFKL framework.
Moreover,  one of the two big parts of radiative corrections beyond the LL
approximation, namely,  the NLL contributions to the two gluon Green's functions 
that link the three tagged jets, do not significantly change the functional behaviour of
$R^{22}_{12}$ on $Y$. The other big part of the beyond the LL corrections would be
the NLO corrections to the jet vertices and these need also to be taken into account. 
In addition, it is crucial to investigate whether  
fixed order calculations and studies with the BFKL inspired 
Monte Carlo  {\bf BFKLex}~\cite{Chachamis:2011rw,Chachamis:2011nz,Chachamis:2012fk,
Chachamis:2012qw,Caporale:2013bva,Chachamis:2015zzp,Chachamis:2015ico,Chachamis:2016ejm} 
give similar results. Predictions from the usual all-purpose collinear
Monte Carlo codes are also needed to complete the picture from the theoretical side.
To conclude with, a dedicated experimental analysis on the proposed ratio observables $R^{MN}_{PQ}$ 
based on  existing (and future) LHC data will answer the question of whether
these observables qualify as new probes for the BFKL dynamics.
\begin{flushleft}
%\vspace{-.3cm}
{\bf \large Acknowledgements}
\end{flushleft}
%\vspace{-.3cm}
GC acknowledges support from the MICINN, Spain, 
under contract FPA2013-44773-P. 
ASV acknowledges support from Spanish Government 
(MICINN (FPA2010-17747,FPA2012-32828)) and, together with FC and FGC, 
to the Spanish MINECO Centro de Excelencia Severo Ochoa Programme 
(SEV-2012-0249). FGC thanks the Instituto de F{\'\i}sica Te{\'o}rica 
(IFT UAM-CSIC) in Madrid for warm hospitality.

\end{document}